\documentclass[onecolumn,showpacs,preprintnumbers,amsmath,amssymb]{revtex4-2}
\usepackage{graphicx}
\usepackage{dcolumn}
\usepackage{bm}
\usepackage{rotating}
\usepackage{lipsum}

    {\par\small\addvspace{4.5ex plus 1ex}%
     \vskip -\parskip
     \ifx\relax#1\relax
        \def\@decl@date{}%
     \else
        \def\@decl@date{\NEWfeature{#1}}%
     \fi
     \noindent\hspace{-\leftmargini}%
     \begin{tabular}{|l|}\hline\ignorespaces}%
    {\\\hline\end{tabular}\nobreak\@decl@date\par\nobreak
     \vspace{2.3ex}\vskip -\parskip}

\def\Red{} 
\def\Black{} 

\def\lpybox#1{} 
\def\lpybox#1{\fbox{\Red{\bf#1}\Black}} 

\def\D{$^2\mathrm{D}_{3/2}$\,}
\def\S{$^2\mathrm{S}_{1/2}$\,}
\def\P{$^2\mathrm{P}_{1/2}$\,}

\begin{document}
\title{Background reduction in $^{136}$Xe double beta decay experiments through direct barium ion detection}
\author{A. Peralta Conde}
\affiliation{Universidad Internacional de la Rioja (UNIR). www.unir.net. Spain.}
\email{alvaro.peralta@unir.net}
\begin{abstract}
Tagging barium ions in double beta decay experiments involving $^{136}$Xe offers a promising pathway to achieving an almost background-free environment, which is essential for addressing key unresolved questions in neutrino physics, such as the nature of neutrinos and their mass hierarchy. In this manuscript, we present a novel detection scheme that relies exclusively on the intrinsic energy levels of the barium ion. This method eliminates the need for additional additives in the xenon vessel, thereby simplifying the experimental setup and enhancing the potential sensitivity of the experiment.
\end{abstract}
%
\maketitle
\section{Introduction}

Understanding neutrinos represents one of the most significant milestones in particle Physics of the 21st century. Despite tremendous experimental and theoretical efforts, two major questions remain unresolved: What is the nature of neutrinos, and what is the ordering of their masses, known as the neutrino mass hierarchy? Obtaining an answer to this questions will provide crucial information about fundamental aspects of our universe, such as the matter-antimatter asymmetry problem, which we strongly believe exists but remains experimentally unproven (see for example \cite{Athar22} and reference therein). Thus, these challenges are not only significant for Particle Physics but also essential for a deeper understanding of other fields in Physics like Cosmology or the fundamental framework of the Standard Model.  

Related to the neutrino mass hierarchy problem, which seeks to elucidate the ordering of neutrino masses, specifically, whether there are two heavier neutrinos and one lighter one or vice versa, several experimental efforts have been undertaken. Key experiments include T2K (Tokai to Kamioka) in Japan (see for example \cite{Abe23}), NOvA in the USA (see for example \cite{Acero22}), and the upcoming Deep Underground Neutrino Experiment (DUNE) \cite{Abed21}. These experiments study neutrino oscillations over long distances, exploiting the fact that neutrinos can change flavor as they propagate through space and matter, with the goal of extracting information about the mass hierarchy. 
 
These large-scale experiments require complex, expensive installations due to the weak interaction of neutrinos with matter. However, a novel approach by Yoshimura et al. from Okayama University, known as Radiative Emission of Neutrino Pair (RENP), proposes a more cost-effective method. RENP uses quantum optics and laser physics to induce collective de-excitation in a medium, leading to the emission of a photon and a neutrino pair (see for example \cite{Yoshimura15}). By analyzing the photon’s energy spectrum, this technique may allow researchers to determine individual neutrino masses and infer the mass hierarchy, and, in principle, it seems also possible to obtain valuable information about the nature of neutrinos.

Regarding the neutrino nature, several scientific collaborations are currently investigating whether neutrinos are Majorana or Dirac particles, that means, whether neutrinos and antineutrinos are the same particle or different ones. Notable experiments include GERDA based on high resolution calorimeters with high purity enriched $^{76}$Ge crystal diodes located at the Laboratori Nazionali del Gran Sasso, Italy \cite{Agostini20}; EXO-200, which uses isotopically enriched liquid $^{136}$Xe at the Waste Isolation Pilot Plant in New Mexico, USA (see for example \cite{Albert16}); KamLAND-Zen in Japan, also focused on $^{136}$Xe (see for example \cite{Abe23}); and NEXT at the Canfranc Underground Laboratory in Spain, employing high-pressure $^{136}$Xe (see for example \cite{Novella23} and references therein). 

Specifically, the NEXT experiment focuses on a very rare nuclear transition in which a nucleus with Z protons decays into a nucleus with Z + 2 protons, while maintaining the same mass number A. In this process, two neutrons are converted into two protons, resulting in the emission of two beta particles and two electron antineutrinos. There are two possible modes for this transition: the standard two-neutrino mode ($\beta\beta2\nu$) and the neutrinoless mode ($\beta\beta0\nu$). In the two-neutrino mode ($\beta\beta2\nu$), two standard beta decays occur, emitting two electrons and two electron antineutrinos ($(Z, A)\rightarrow(Z+2, A)+2e^-+2{\bar{\nu}}_e$). This mode has a typical half-life in the range of $10^{18}-10^{21}$ years and has been observed in different experiments (see for example the results of the KamLAND-Zen collaboration, \cite{Gando13} and references therein). Conversely, the neutrinoless mode ($\beta\beta0\nu$) take place without the emission of neutrinos  ($(Z, A)\rightarrow(Z+2, A)+2e^-$) and has a predicted half-life longer than $10^{26}$ years \cite{Gando16}, significantly longer than the two neutrino mode. This transition, despite the experimental efforts, has not been observed so far. The existence of this mode violates the lepton-number conservation and can only occur if neutrinos are Majorana particles \cite{Majorana37}. If neutrinos are confirmed to be Majorana particles, it would provide a mechanism for leptogenesis and a potential explanation for the matter-antimmater assymetry in the early instants of the Universe \cite{Sakharov67, Fukugita86, Mohapatra80}.

From an experimental point of view, the search for the $\beta\beta0\nu$ transition represents a genuine tour de force. The extraordinarily long half-life associated with this process requires the combination of extremely large detection volumes with long exposure times to have any hope of capturing a clear, unambiguous signal. Moreover, given the anticipated weakness of the $\beta\beta0\nu$ signal, it becomes crucial to minimize any background noise that could lead to false positives. To achieve this, all $\beta\beta$ decay experiments are constructed using ultra-pure materials and operate in deep underground facilities to reduce the impact of cosmic rays and other environmental radiation. A major source of background noise, however, arises from the standard two-neutrino double-beta decay ($\beta\beta2\nu$), which has a much shorter half-life compared to the neutrinoless mode. Despite this, detectors with excellent energy resolution can differentiate between the two modes, as the emitted electrons from $\beta\beta0\nu$ carry a distinct energy signature compared to those from $\beta\beta2\nu$ \cite{Elliot02}. Nevertheless, while these strategies significantly reduce background noise, a certain level of residual noise is inevitable. As a result, sophisticated data analysis techniques are necessary to extract the faint neutrinoless signal from the background, ensuring that no false positives are detected and improving the sensitivity of the experiment.

In the NEXT experiment, the searched decay processes are $^{136}$Xe $\rightarrow^{136}$Ba$^{+2}+2e^-+2\bar{\nu}$ (two-neutrino mode) and $^{136}$Xe $\rightarrow^{136}$Ba$^{+2}+2e^-$ (neutrinoless mode). One of the most promising strategies to reduce the background noise is to correlate the detection of the emitted electrons with the detection of the $^{136}$Ba atom produced in the nuclear transition. This is because no known radioactive process in xenon can generate such a signal, i.e., the simultaneous detection of electrons and barium atoms, except for $\beta\beta$ transitions. This concept is referred to as "Barium Tagging" (BaTa), and if implemented successfully, it would increase the sensitivity of $\beta\beta$ experiments by several orders of magnitude. Several approaches for BaTa have been proposed so far. The earliest idea for implementing BaTa in a xenon time-projection chamber (TPC) dates back to 1991, proposed by M. K. Moe \cite{Moe91}, and has been extensively studied since then \cite{Danilov00, Sinclair11, Mong15}. Additionally, the nEXO collaboration has made significant advances in identifying barium atoms in liquid xenon. In contrast to xenon gas, for liquid xenon recombination is quite likely being the Ba atoms distributed across charge states from 0 to +2 \cite{nEXO19, Albert15}. For high-pressure gaseous xenon experiments, another approach is to use a specially designed molecule that captures the Ba$^{+2}$ ion, which then undergoes a change in its fluorescence properties. By illuminating this molecule with lasers and analyzing the emitted fluorescence spectrum, it may be possible to confirm the capture of a barium atom. Various theoretical and experimental efforts have been made to develop this approach, including work by D. R. Nygren, P. Thapa, and others (see for example \cite{Nygren15, Thapa19}), as well as more recent developments in this area \cite{Rivilla20}.

In this manuscript, we propose a novel and simplified approach to BaTa for the NEXT experiment, focusing on the direct detection of fluorescence from excited barium ions. Specifically, we aim to detect the characteristic fluorescence emitted when barium ions, produced during the $\beta\beta$ nuclear transition of $^{136}$Xe, are excited by laser radiation. Unlike previous methods that rely on specially designed molecules to capture and tag barium atoms, our approach eliminates the need for molecular tagging agents, significantly simplifying the experimental setup. This method requires only precise laser excitation of the barium ions, followed by the detection of their fluorescence, which can be distinguished by its unique spectral signature. By reducing complexity, this approach not only streamlines the detection process but also holds potential for enhancing the sensitivity and scalability of future BaTa implementations.

\section{Barium Tagging}
As it was discussed above, the NEXT experiment relies on the detection of the doubly beta decay neutrinoless desintegration of $^{136}$Xe atoms to produce $^{136}$Barium doubly ionized atoms plus two electrons either accompanied by the emission of two neutrinos (two-neutrino mode) or without neutrino emission (neutrinoless mode). The discrimination between both modes is realized in terms of the emitted electron energies. In the neutrinoless process, a very well-defined energy for the electrons is expected, close to the Q-value of the reaction, while in the neutrino process, the electron energy exhibits a broad distribution, as part of the energy is carried away by the neutrinos. Also, in the experiment, a full reconstruction of the trajectories of the emitted electrons is expected, tracking their paths from the decay point within the detection chamber to the readout plane (see, for example, \cite{Haefner24} and references therein). This reconstruction algorithm will allow to pinpoint the likely location of the decay with a high degree of precision. This result is critical for the search for the resulting $^{136}$Ba near the decay site. Since electrons travel much faster than the barium ions, the latter can be considered essentially stationary ("frozen") during the time it takes for the electron signal to be detected.

Tagging doubly ionized barium ions  $^{136}$Ba$^{++}$ directly via laser-induced fluorescence presents significant challenges. The first excited state of Ba$^{++}$ lies at 132770.79\,cm$^{-1}$, corresponding to an energy of 16.46\,eV, which translates to extreme ultraviolet (XUV) radiation at a wavelength of about 75\,nm. Generating such radiation in a laboratory setting is exceedingly difficult due to its absorption by air and the lack of readily available optics for XUV wavelengths. To overcome this difficulty, we rely on the assumption that the Ba$^{++}$ ions will undergo charge exchange with the Xe atoms to form Ba$^{+}$, whose excited states lie in the visible part of the spectrum. This charge exchange process is favored by the high pressure of Xe gas in the NEXT vessel, approximately 10\,atm, and the significant difference in ionization energies between Ba$^{++}$ (35.84\,eV) and Xe (12.13\,eV) \cite{Sanso10}. Thus, xenon atoms can donate an electron to the Ba$^{++}$ ions via collisions, allowing for a transition from Ba$^{++}$ to Ba$^+$. Furthermore, at such high pressure it is expected the formation of weakly bounds van der Waals molecules between the barium ions and the neutral xenon \cite{Abde13}. In these molecules, there is a weak interaction between the electronic clouds which promotes the transfer of an electron to the Ba$^{++}$ ions. However, we do not expect the absorption of a second electron by Ba$^{+}$ given the difference in ionization energy between Ba$^{+}$ ions, 10\,eV, and Xe atoms, 12.13\,eV, \cite{Curry04} which makes such a process energetically unfavorable.

It is straightforward to make a rough estimation of the number of collisions in the NEXT vessel. We assume that Xe atoms and Ba$^++$ ions behave like solid spheres with diameters d$_1$ and d$_2$ respectively, and occupy a volume V. After the nuclear transition, the velocity of the Ba$^++$ ions is almost zero, and in any case negligible when compared with the velocity of the Xe atoms v$_1$. According to this, for a collision to between a Ba ion and a Xe atom to occur, a Xe particle must be inside the cylindrical volume defined by:
\begin{equation}
\mathbf{V}_{\mathbf{Cyl}}=\mathbf{<v_{12}> dt \, \pi \left(\frac{d_1+d_2}{2}\right)^2=<v_1> dt \, \pi \left(\frac{d_1+d_2}{2}\right)^2}.
\end{equation} 
Defining the collision diameter as 
\begin{equation}
\mathbf{d_{12}=\frac{d_1+d_2}{2}},
\end{equation}
the number of Xe particles N inside the volume $\mathbf{V_{Cyl}}$ (being N$_1$ the total number of Xe atoms) is defined by
\begin{equation}
\mathbf{N=\frac{N_1}{V}V_{Cyl}=\pi d_{12}^2 <v_{1}> dt \frac{N_1}{V}},
\end{equation}
and therefore the number of collisions per unit of time
\begin{equation}
\label{collt}
\mathbf{z=\pi d_{12}^2 <v_{1}>\frac{N_1}{V}}.
\end{equation}
Assuming thermodynamic equilibrium, and that both gases behave as ideal gases, we have
\begin{equation}
\mathbf{<v_1>=\left(\frac{8KT}{\pi m_1}\right)^{1/2}}
\end{equation}

\begin{equation}
\mathbf{PN_{A}=\frac{N_1}{V}RT},
\end{equation}

being \textbf{K} the Boltzmann constant, \textbf{T} the temperature, \textbf{V} the volume of the vessel of NEXT, N$\mathbf{_A}$ the Avogadro constant, and \textbf{R} the ideal gas constant. Substituting these expressions in Eq.\,\ref{collt} we finally obtain

\begin{equation}
\label{collt_2}
\mathbf{z=\pi d_{12}^2  \frac{PN_A}{R T}\left(\frac{8KT}{\pi m_1}\right)^{1/2}}.
\end{equation}

According to the documentation, the Xe pressure in the NEXT vessel will be of around 10\,atm. Assuming a room temperature of 298\,K, and a diameter for both Xe atoms and Ba$^++$ ions of approximately 2.5$\mathbf{\AA}$, the collision rate can be estimated to be approximately:
\begin{equation}
\label{coll_rate}
\mathbf{z=10^{10}\,s^{-1}}.
\end{equation}
This high collision rate is sufficient to ensure the transition Ba$^{++}\rightarrow$Ba$^{+}$.

\section{Barium tagging with singly ionized barium ions}
\begin{figure}[ht!]
\begin{center}
\includegraphics[width=8.3cm, height=5cm]{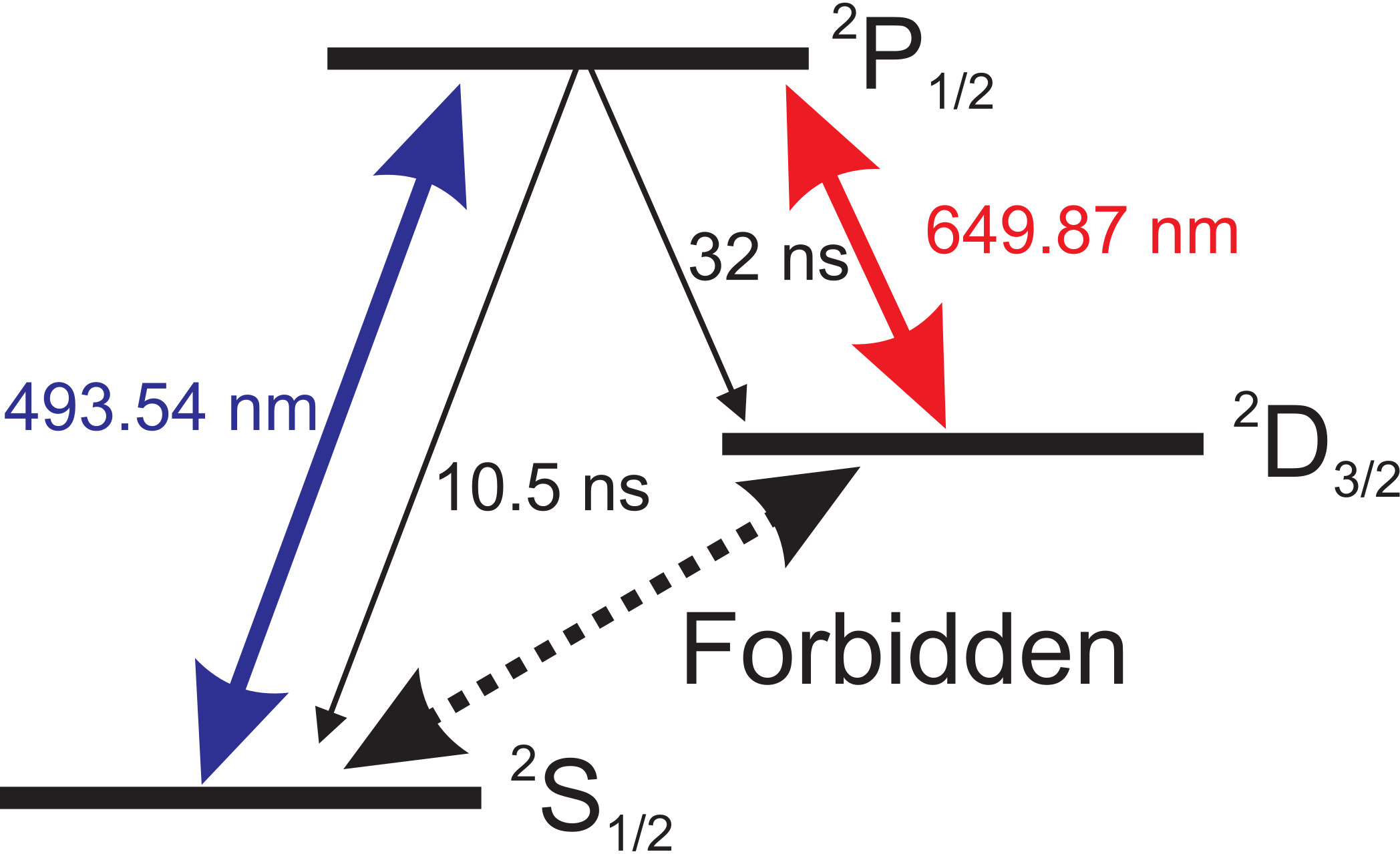}
\caption{\label{levelscheme} Level scheme of the three lowest states of Ba$^+$ ions and the corresponding radiative lifetimes. The notation corresponds to $^{2\mathrm{S}+1}\mathrm{L}_\mathrm{J}$ being S the spin angular momentum, L the angular momentum, and J=L+S.}
\end{center}
\end{figure}

Figure,\ref{levelscheme} shows the first three states of Ba$^+$ ions along with their corresponding radiative lifetimes. It is plausible to consider that barium ions may appear after the disintegration of xenon in the ground state. The level structure of Ba$^+$ is somewhat peculiar, as the first excited state, specifically the state $^2\mathrm{D}_{3/2}$, is not optically accessible from the ground state when considering the dipole approximation and single-photon transitions \cite{Curry04}. It is worth remembering the selection rules for $\pi$ polarized light for single photon absorption:
\begin{equation}
\Delta\mathrm{L}=1\hspace{1cm}\Delta\mathrm{S}=0 \hspace{1cm} \Delta\mathrm{J}=0,\pm1;\,\Delta m_\mathrm{J}=0\hspace{0,5cm} (\text{with forbidden}\, \Delta\mathrm{J}=0;\,\mathrm{m_J}=0\rightarrow\mathrm{m'_J}=0)
\end{equation}

\begin{figure}[ht!]
\begin{center}
\includegraphics[width=9cm, height=5.7cm]{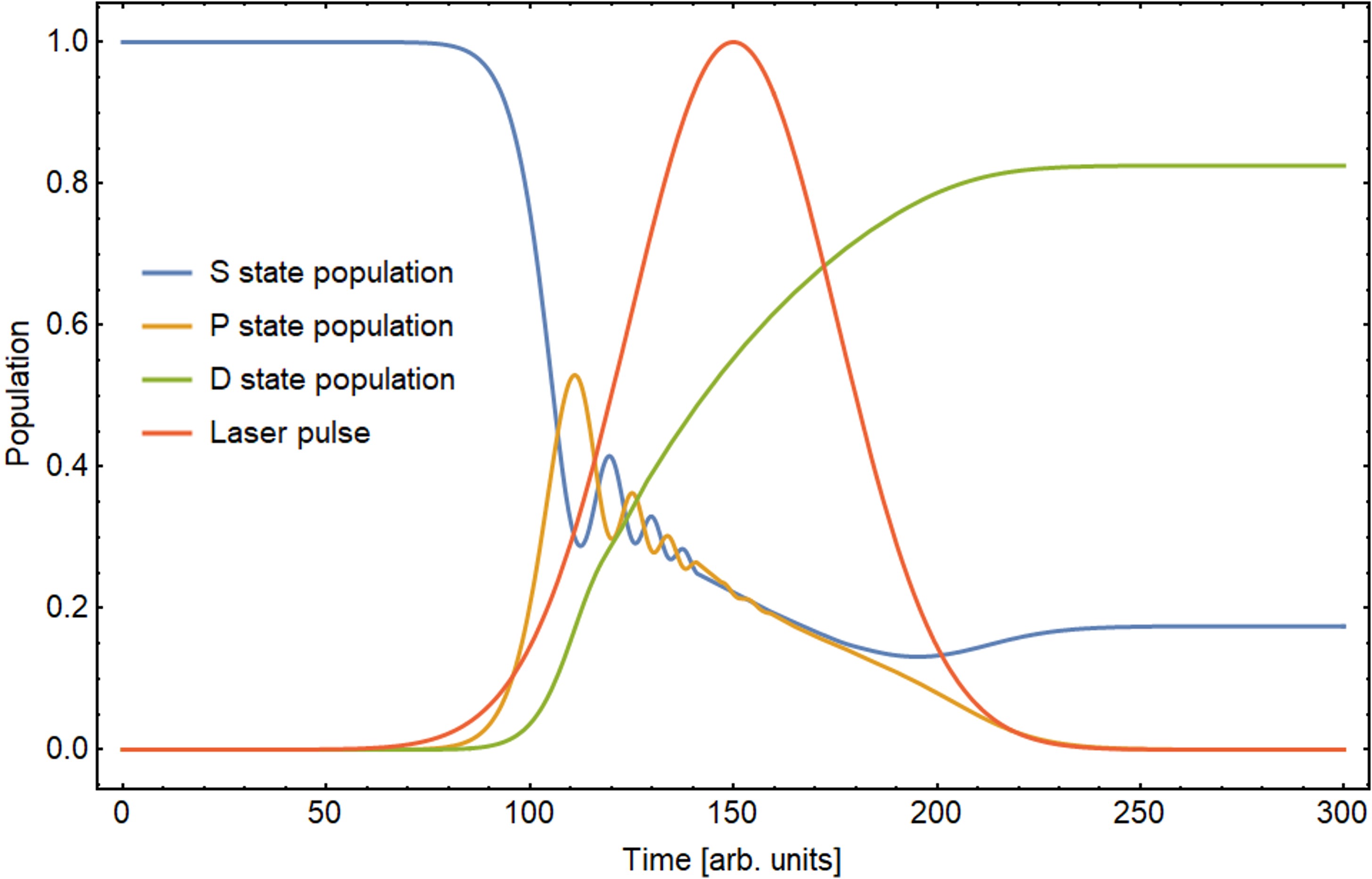}
\caption{\label{grafico1} Population dynamics following the laser pulse excitation of the \S $\rightarrow$ \P transition.}
\end{center}
\end{figure}

According to the previous discussion, and considering that the first excited state of Ba$^+$ ions is a metastable state, the proposed level scheme for the implementation of BaTa involves resonant excitation of the \S$\rightarrow$\P transition, followed by the collection of fluorescence emitted during the \P$\rightarrow$\D and/or \P$\rightarrow$\S transitions (see Fig.,\ref{levelscheme}). A significant drawback of this scheme, inherent to the level structure of Ba$^+$ ions, is that every time an infrared (IR) photon is emitted during the \P$\rightarrow$\D transition, the population becomes trapped in the metastable \D state. This phenomenon is illustrated in Fig.,\ref{grafico1}, which depicts the population dynamics after the \S$\rightarrow$\P transition is excited by a laser pulse. As shown, once the population is excited to the \P state, it can spontaneously decay to both the \S and \D states, emitting fluorescent photons that could, in principle, be detected. However, once the population reaches the \D state, it remains trapped, hindering further transitions and fluorescence detection.

While one might consider that numerous collisions with Xe atoms could induce transitions from the metastable \D state to the ground state of the Ba ions, this scenario seems unlikely due to the significant energy difference between the two systems. The ionization potential of Xe is 12.13\,eV, whereas the energy difference for the radiatively forbidden transition \D$\rightarrow$\S is only 0.6\,eV. This disparity significantly reduces the likelihood of such a transition occurring, making it impractical for experimental purposes, where a sufficient number of detected photons is required to ensure a positive identification of the Ba atom and, hence, of a double neutrino transition. To achieve sufficient statistical significance, it is therefore crucial to develop a mechanism to repump the population trapped in the metastable \D state back to either the \P state or the \S state. Potential strategies could include using an additional laser pulse. We will discuss possible experimental implementations to address this issue in the following section.

\section{Population Dynamics}
According to the previous discussion, the optimal approach for BaTa involves utilizing two lasers: one laser to couple the \S and \P states, and a second laser to couple the \P and \D states. This configuration is commonly referred to as a lambda system in the quantum optics literature. The population dynamics of this system is governed by the time-dependent Schr\"{o}dinger equation: 
\begin{equation}
\label{timedepsch} 
i\hbar\frac{\partial\Psi(t)}{\partial t}=H(t)\Psi(t),
\end{equation}
where $\Psi(t)$ represents the statevector of the system, and $H(t)$ the Hamiltonian that incorporates the interaction with a radiation field. 

If we consider the Rotating Wave Approximation, the Hamiltonian can be significantly simplified. When considering the Rotating Wave Approximation (RWA), the Hamiltonian can be significantly simplified. This approximation is valid under the assumption that our primary interest lies in the population dynamics over the temporal variation of the laser's envelope, rather than in studying the population dynamics over multiple optical cycles.  Consequently, the rapid oscillations inherent in the laser field can be averaged out. The RWA provides a reliable approximation as long as the laser pulse contains several optical cycles. However, for ultrashort laser pulses comprising only one or two cycles, the RWA becomes inapplicable \cite{Sh90}. Under the RWA, the Hamiltonian is expressed as:

\begin{equation}
\label{Hamilt}
\rm H=\frac{\hbar}{2}\left(
\begin{array}{ccc}
0 & \rm \Omega_P  & 0\\
\rm \Omega_P  & \rm 2\Delta_P & \rm \Omega_S  \\
\rm 0 & \rm \Omega_S  &\rm 2(\Delta_P-\Delta_S)
\end{array}
\right).
\end{equation}
The off diagonal terms in Eq.\,\ref{Hamilt} are the Rabi frequencies, i.e., the interaction energy divided by $\hbar$, 
\begin{equation}
\label{ch2definitonofrabi}
\Omega(t)=-(\textbf{d}\cdot\textbf{e})\mathcal{E}(t)/\hbar, 
\end{equation}
where $\textbf{d}$ represents the dipole moment of the transition. For moderate laser intensities, the contributions from the magnetic dipole interaction, as well as higher-order interaction terms, can be neglected. The vector $\textbf{e}$ denotes the polarization direction of the laser, while $\mathcal{E}(t)$ represents the slow varying envelope of the laser electric field. The diagonal terms in the Hamiltonian correspond to the laser detunings from resonance.

In order to account for losses, it is more effective to work within the density matrix formalism, wherein the Liouville–von Neumann equation governs the population dynamics of the system. The evolution of the density matrix $\rho$ is describer by the equation
\begin{equation}\label{Liouville}
\rm \imath \hbar \frac{d\rho}{d t}=\left[H, \rho \right]
\end{equation}
being $\rho$ the density matrix given by
\begin{equation}
\rm \rho=\left(
 \begin{array}{ccc}
\rho_{11} & \rho_{12} & \rho_{13} \\
\rho_{21} & \rho_{22} & \rho_{23} \\
\rho_{31} & \rho_{32} & \rho_{33}
 \end{array}
\right).
\end{equation}
In this representation, the diagonal terms $\rm \rho_{ii}$ represent the populations of the state $\rm |i\rangle$, while the non-diagonal elements correspond to the coherences satisfying the relation $\rho_{ij}=\rho^*_{ji}$. 
The Hamiltonian H is defined as per Eq.\,\label{Hamilt}. 

Since in this formalism the observables are directly the populations rather than the probability amplitudes like in the time-dependent Schr\"{o}dinger equation, the introduction of the radiative losses, spontaneous emission, is straighforward. The relevant equation then reads (see for example \cite{Shore08}):
\begin{equation}
\dfrac{d}{dt}\rho_{mn}(t)=-i\sum_k\left[H_{mk}(t)\rho_{kn}(t)-\rho_{mk}(t)H_{kn}(t)\right]-\rho_{mn}(t)\sum_k\frac{1}{2}\left(A_{mk}+A_{nk}\right)+\delta_{mn}\sum_k\rho_{kk}(t)A_{km},
\end{equation}
where $H_{j}$ correspond to the element $i,j$ of the Hamiltonian defined in Eq.\,\ref{Hamilt}, and $A_{mk}$ are the spontaneous decay rates. Notably, in the density matrix formalism, observables are directly related to the populations, as opposed to the probability amplitudes used in the time-dependent Schrödinger equation. This facilitates a straightforward incorporation of losses such as radiative losses like spontaneous emission. The equations describing the time evolution of the density matrix elements are given by:
\begin{equation}
 \frac{d}{dt}\rho_{mn}(t) = -i \sum_k \left[ H_{mk}(t) \rho_{kn}(t) - \rho_{mk}(t) H_{kn}(t) \right] - \rho_{mn}(t) \sum_k \frac{1}{2} \left( A_{mk} + A_{nk} \right) + \delta_{mn} \sum_k \rho_{kk}(t) A_{km}. \end{equation}

For a successful experimental implementation of BaTa, it is essential to analyze the interaction dynamics based on key experimental laser parameters, such as pulse duration, laser intensity, or laser detuning. According to previous definitions, it is straightforward to show that the solution to the time-dependent Schrödinger equation, or the Liouville–von Neumann equation, for a two-state system on resonance, i.e., $\Delta=0$, can be written as:
\begin{equation}
\label{Population_equations}
\rm P_1(t)=\cos^2A(t) \qquad P_2(t)=\sin^2A(t)
\end{equation}
where $\rm A(t)$ is the pulse area, defined by
\begin{equation}
A(t)=\int_{-\infty}^t\Omega(t')dt'.
\end{equation}
According to Eq.\ref{Population_equations}, for a strong interaction, meaning the maximization of the transition probability between ground and excited states, saturation of the transition is required. This condition implies inducing Rabi oscillations which necessitates $\Omega \tau\gg 1$, being $\tau$ the pulse duration \cite{Sh90}. Since the Rabi frequency $\Omega$ can be express in terms of the laser intensity as: 
\begin{equation}
\label{rabiI}
\Omega=\frac{\mu}{\hbar}\sqrt{\frac{2I}{c\epsilon_0}}
\end{equation}
and the laser intensity I can be expressed as:
\begin{equation}
I=\frac{E_{pp}}{S\tau}
\end{equation}
where $E_{PP}$ is the energy per pulse and S is the effective interaction area, the saturation condition scales with $\sqrt{\tau}$ as follows:
\begin{equation}
\Omega\tau\sim(1/\sqrt{\tau})\tau=\sqrt{\tau}.
\end{equation}
This dependence of the saturation condition on the laser pulse duration reveals that the use of ultrashort laser pulses (e.g., short picosecond or femtosecond pulses) is unfavorable for achieving significant population transfer. In fact, with ultrashort lasers, reaching the saturation regime requires increasing laser intensities substantially, which can induce unwanted nonlinear processes, such as Stark shifts, multiphoton absorption via virtual states or even tunneling ionizations. Thus, for effective population transfer, nanosecond or long picosecond lasers are more suitable, as they allow sufficient pulse duration to reach saturation without necessitating excessively high intensities. Furthermore, the use of nanosecond lasers simplifies the experimental implementation, as their relatively narrow bandwidth minimizes distortions during air propagation in the visible spectrum. Additionally, nanosecond pulses require no specialized optics, making their manipulation and generation straightforward and cost-effective in laboratory settings.

\section{Numerical simulations}
According to the discussion in Fig.\,\ref{grafico1}, to achieve a sufficient fluorescence signal for detecting the single Ba atom produced in the nuclear $\beta\beta$ transition, it is necessary to repump the population trapped in the \D  metastable state following radiative decay after excitation by a laser, referred to as the Pump laser.  This repumping is achieved by introducing a second laser, known as the Stokes laser, forming a pump/repump cycle. This pump/repump scheme together with a tracking algorithm of the Ba ion, ensures continuous cycling of the population, enabling the collection of fluorescence from both decay channels. 

Fluorescence detection can occur while the lasers are on, as fluorescence emission is isotropic, while the laser beams are directional. However, this approach may be compromised by unavoidable scattering from both the Pump and Stokes lasers in the high-pressure Xe gas environment or at the entrance window. A more effective approach may be to wait until the lasers are turned off before initiating fluorescence detection, i.e., to pulse the fluorescence detection. This strategy may minimizes interference from scattered laser light, enhancing the signal-to-noise ratio.

\begin{figure}[ht!]
\begin{center}
\includegraphics[width=9cm, height=9cm]{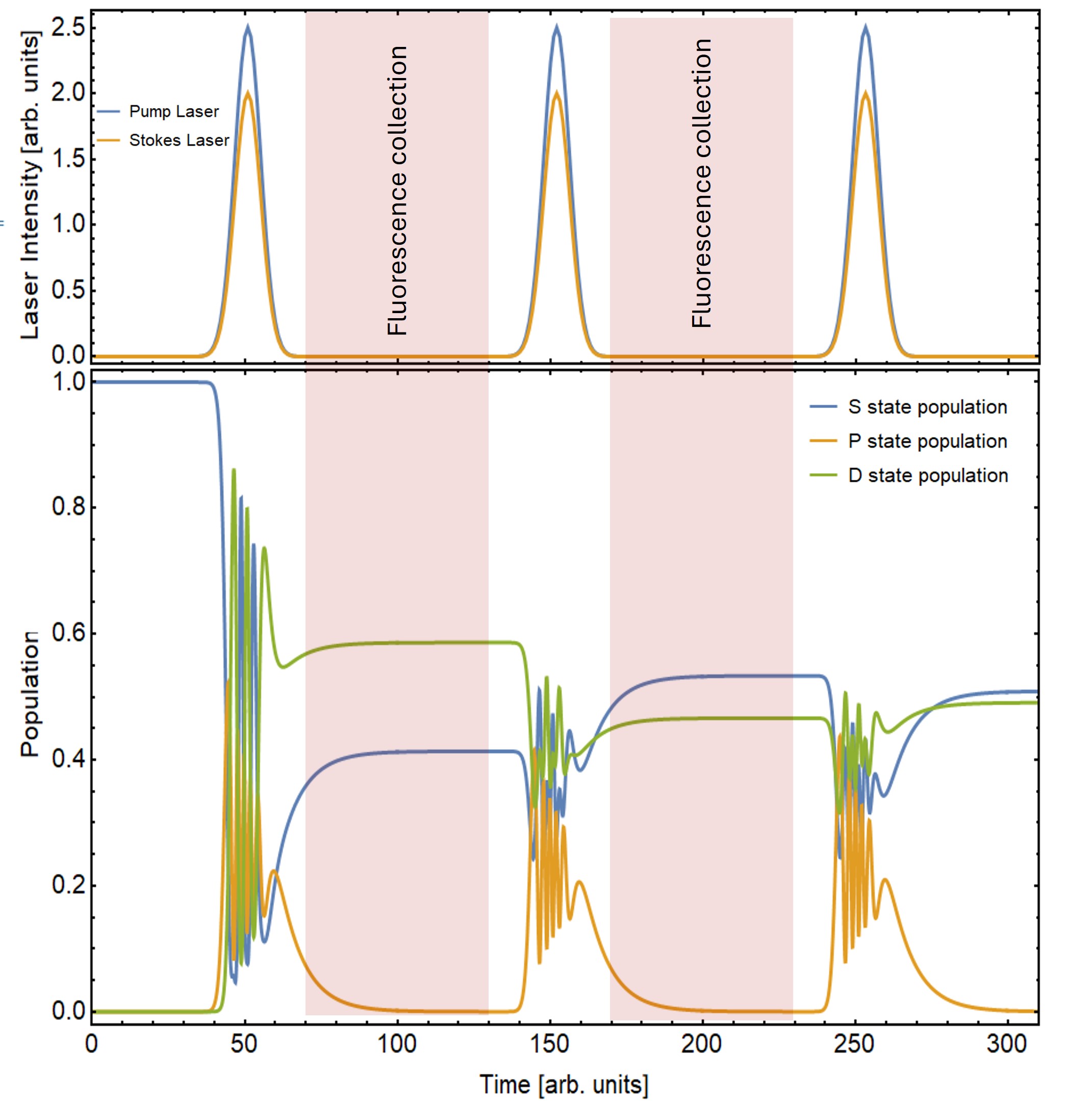}
\caption{\label{grafico2} Upper panel: Pump and Stoke lasers. Lower panel: Population dynamics. The fluorescence detection window is also indicated.}
\end{center}
\end{figure}

To optimize this scheme, it is essential to control the experimental parameters to maximize the population of the \P state once the lasers are turned off. Figure\,\ref{grafico2} shows the laser excitation and repumping cycles scheme and the population dynamics alongside the designated detection window. Fine-tuning parameters such as pulse timing, laser power, and pulse duration will be crucial for maximizing fluorescence yield and achieving reliable Ba atom detection in this setup.

\begin{figure}[ht!]
\begin{center}
\includegraphics[width=9cm, height=9cm]{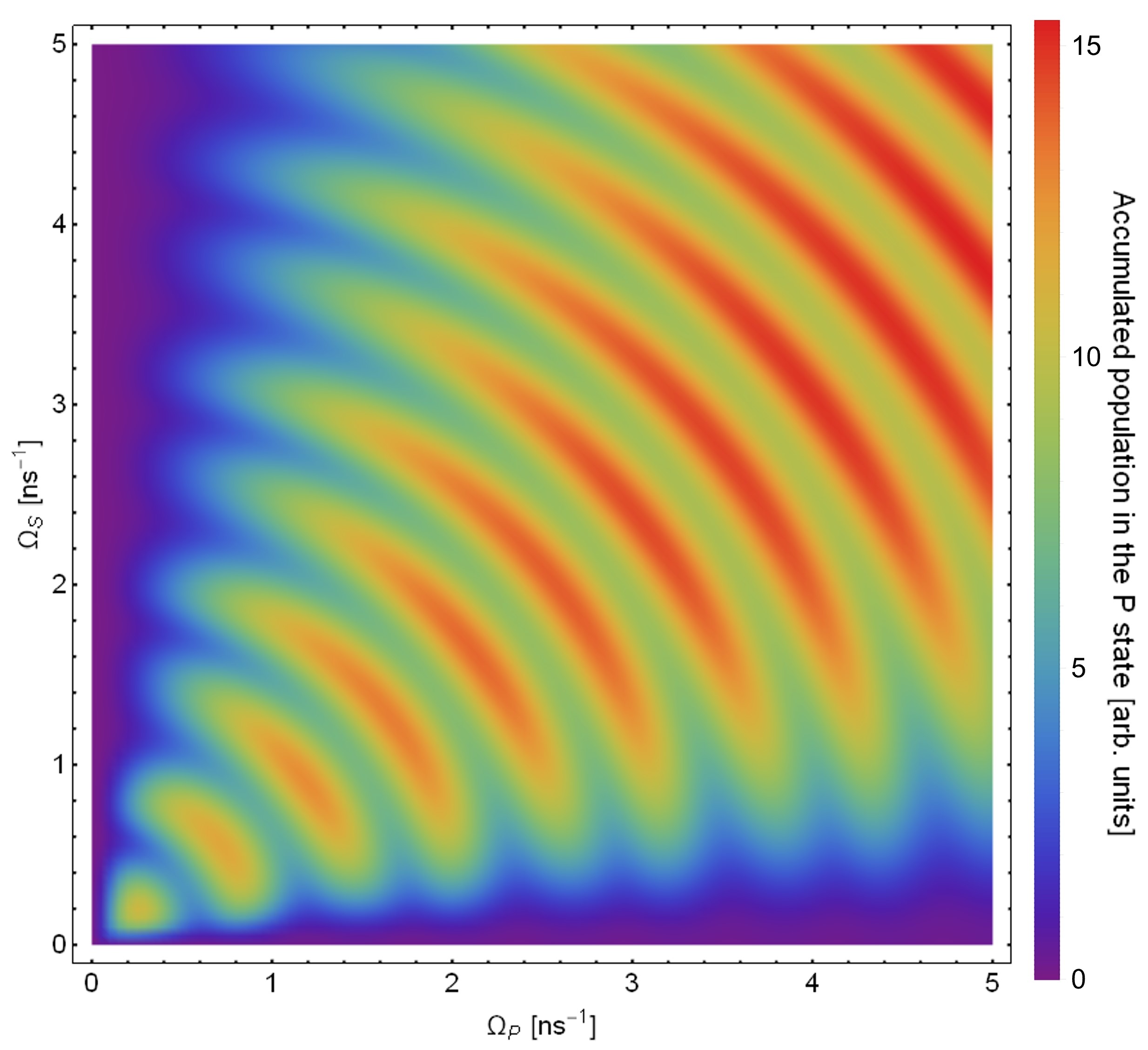}
\caption{\label{Simu2} Accumulated population in the \P state after 200 repump cycles as a function of Pump and Stokes lasers Rabi frequencies. The lifetime of the radiative transitions \P$\rightarrow$\S and \P$\rightarrow$\D is 10.5 and 32\,ns respectively. The laser pulses are coincident and with a pulse duration of 10\,ns.}
\end{center}
\end{figure}

Figure \ref{Simu2} illustrates the accumulated population in the \P state over 200 pump/repump cycles. This accumulated population represents the sum of the \P state population after each cycle, which is directly proportional to the fluorescence signal, plotted as a function of the Rabi frequencies of the Pump and Stokes lasers. It should be noted that in the simulation, the initial conditions of each cycle are defined as the final conditions from the preceding cycle, ensuring an accurate reflection of continuous dynamics. Additionally, the integration time is set so that no residual population remains in the \P state before the next pump/repump cycle begins. 

The simulations use Rabi frequencies as primary variables rather than laser intensities because the Rabi frequency directly characterizes the strenght of the laser-matter interaction. The relationship between Rabi frequency and laser intensity depends on the dipole transition moment, which varies across systems. Specifically, the Rabi frequency is defined by Eq.,\ref{ch2definitonofrabi}:
\begin{equation}
\Omega(t)=-(\textbf{d}\cdot\textbf{e})\mathcal{E}(t)/\hbar.
\end{equation}
 The relation between the laser intensity and the laser electric field is given by
 \begin{equation}
I=\frac{1}{2}c \epsilon_0 c \mathcal{E}^2
\end{equation}
The transition dipole moment $\textbf{d}$ is directly related with the radiative lifetime $\tau_{ij}$ by \cite{Hilborn}:
\begin{equation}
\text{d}_{ij}^2=\frac{3\epsilon_0h\lambda_{ij}^3}{16\pi^3}\frac{1}{\tau_{ij}}
\end{equation}
being $\lambda_{ij}$ the excitation wavelength of the transition. 
For barium, an alkaline earth metal, the transition dipole moments are relatively large, on the order of a Debye (D, 1\,D=3.33\,10$^{-33}$\,Cm) which facilitates experimental implementation.

From Figure\,\ref{Simu2}, we observe that the fluorescence signal increases in the direction of increasing Rabi frequencies with a slight deviation due to the different decay rates of the \P state. This is explained quantitatively in Appendix \ref{AppIncohLimit}. Additionally, the observed oscillations result from the coherent treatment of population dynamics, specifically Rabi oscillations. However, under experimental conditions, where precise control over the different parameters is challenging, these oscillations would likely average out. For example, the Gaussian spatial profile of the laser beam results in a Gaussian distribution of Rabi frequencies across the interaction region, rather than a  flat-top distribution. Thus, it is encouraging that the population in the \P state remains relatively stable over a range of Rabi frequencies, indicating a robustness to unavoidable fluctuations in laser intensity. 

\begin{figure}[ht!]
\begin{center}
\includegraphics[width=18cm, height=8.4cm]{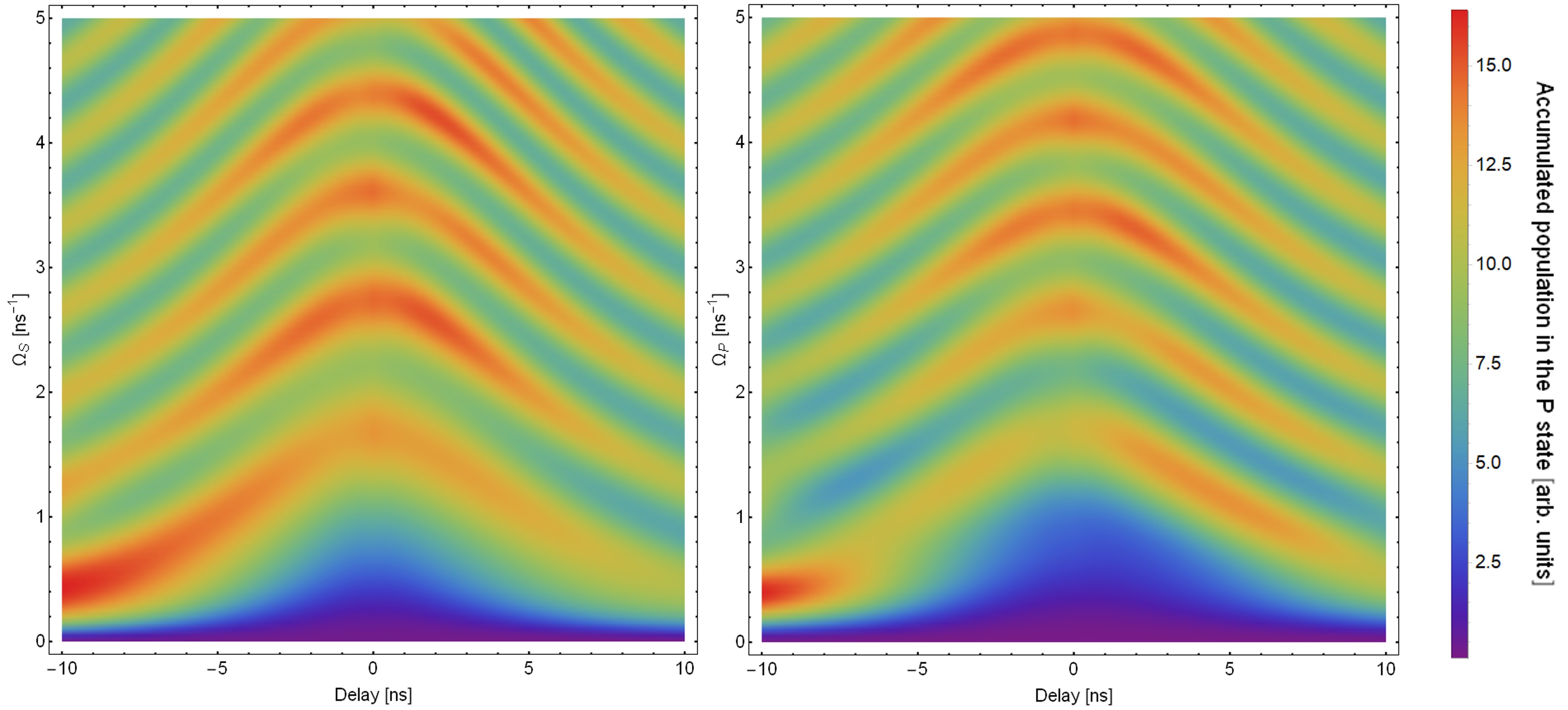}
\caption{\label{Simu3} Accumulated population in the \P state after 200 repump cycles as a function of the Rabi frequency of the Stokes laser (left panel) and Pump laser (right panel), as well as the delay of the Stokes laser with respect to the Pump laser. The simulation data correspond to pulse durations for Pump and Stokes lasers of 10\,ns. In the left panel the Pump Rabi frequency is fixed at $\Omega_P=3.5\,ns^{-1}$, while in the right panel the Stokes Rabi frequency is set to $\Omega_S=2.8\,ns^{-1}$.}
\end{center}
\end{figure}

An important aspect to consider for a potential experimental implementation is the delay between the laser pulses. When working with nanosecond lasers, this parameter is manageable since 1 ns corresponds to a distance of approximately 30 cm. However, if the maximum population in state \P shows a strong dependence on the delay, it could complicate the experimental realization. Figure \ref{Simu3} displays the population of state \P after 200 repumping cycles as a function of the delay, as well as the Rabi frequencies of the Stokes and Pump lasers (left and right panels, respectively). It can be observed that the population of state \P does not exhibit significant dependence on the delay between the Pump and Stokes laser pulses. This finding is promising for experimental implementation, as it suggests robustness against variations in pulse timing. Interestingly, despite the apparent symmetry around zero delay, there is a region, particularly at low Rabi frequencies and long delays, where the population of state \P deviates notably from this simmetry. We attribute this behavior to a coherent phenomenon like for example STIRAP (Stimulated Raman Adiabatic Passage) \cite{Shore08}. These coherent effects naturally emerge as we solve the population dynamics using the density matrix formalism, which inherently accounts for coherent interactions. That said, in an experimental setting, the recorded signal is typically an accumulation over multiple measurement cycles. Consequently, coherent effects may be obscured due to fluctuations in experimental parameters unless the conditions are meticulously controlled. Given the purpose of this study -primarily to demonstrate the detection of Ba atoms- it would be unnecessarily complex to address such coherent effects in an already challenging experimental setup.

To conclude this review of the different experimental conditions that must be taken into account for the implementation of BaTA, we must address the different mechanisms that can either broaden the absorption linewidth or even shift the absorption frequency. In fact, the absorption or emission spectra of radiation are never strictly monochromatic. Since an excited atom or molecule has a certain probability of spontaneous decay to a lower energy state, there is an inherent uncertainty in energy due to the energy-time uncertainty principle, known as the natural linewidth. For atoms, where the lifetimes of excited states typically range on the nanosecond scale, the natural linewidth is of the same order as the linewidth of nanosecond lasers. As a result, nanosecond lasers are the best systems to maximize population transfers. 

Other phenomena can contribute to the broadening of the absorption linewidth. One such phenomenon is Doppler broadening, which arises from the thermal motion of particles relative to the radiation source. In a gas at non-zero temperature, the distribution of particle velocities follows a Maxwell-Boltzmann distribution, meaning that a range of Doppler shifts occurs, leading to an overall broadening of the absorption lines which increases with temperature. Another significant effect is pressure broadening, also known as collisional broadening. Collisions may alter the energy states of the particles involved resulting in a broadening and/or shift of the emission lines. The broadening of the absorption spectrum results in a wider range of resonance frequencies, but at the cost of a reduction in the maximum absorption intensity \cite{Demtroder03}. 

To calculate the Doppler broadening in this context, as a first approximation we can assume that the barium ion produced in the double beta decay rapidly thermalizes with the xenon atoms within the chamber, which is maintained at a temperature of 300 K. The Doppler broadening can be calculated using the following expression:
\begin{equation}
\Delta \nu_D = \frac{2}{\lambda} \sqrt{\frac{2 k_B T \ln 2}{m}}
\end{equation}
where $\lambda$ is the wavelength of the transition, m is the mass of the particle, and $k_B$ the Boltzmann constant. For the \S$\rightarrow$\P this Doppler broadening results in a linewidth of $\Delta\nu_D\simeq0.6$\,GHz.

A more significant consideration for the experimental implementation is the Doppler shift. The @NEXT experiment employs a time projection chamber (TPC), which utilizes a combination of high electric and magnetic fields to perform a three-dimensional reconstruction of particle trajectories, thereby determining the spatial origin of double beta decay events (see for example \cite{Martin16}). In this setup, the Ba ion is projected toward the cathode of the TPC at high velocities, encountering numerous collisions with the surrounding Xe atoms. The Doppler shift can be quantified using the following expression:
\begin{equation}
\Delta\nu_{\text{Doppler}} = \overrightarrow{k} \cdot \overrightarrow{v} = \pm \frac{2\pi v \cos \alpha}{\lambda},
\end{equation}
where  is the velocity of the particle and  is the angle between the velocity vector and the direction of propagation of the radiation.

In the limiting case where collisions are neglected, typical velocities of Ba ions in TPC chambers can reach the order of tens of thousands of meters per second. For counter-propagating incidence, $\alpha=0\,\text{or}\,\pi$, the induced Doppler shift can be estimated as:
\begin{equation}
\Delta\nu_{\text{Doppler}} \simeq 100\,\text{ns}^{-1}.
\end{equation}
This value is significantly larger than the laser linewidth, defined as $\Delta\nu=1/\tau$ , where $\tau$ represents the pulse duration. For laser pulses on the order of tens of nanoseconds, the bandwidth is approximately 100\,$\mu\text{s}^{-1}$. According to these calculations, while it is true that ion mobility can be significantly reduced in the NEXT experiment due to the high-pressure conditions that promote frequent collisions, the Doppler shift still presents a considerable challenge.  Based on this, for successful implementation of BaTa, it is strongly recommended to irradiate the particles from a normal or quasi-normal direction to minimize the Doppler shift. This approach helps ensure that the laser frequency remains well-aligned with the absorption profile of the Ba ions, facilitating more effective resonance and detection.

\begin{figure}[ht!]
\begin{center}
\includegraphics[width=9cm, height=7.8cm]{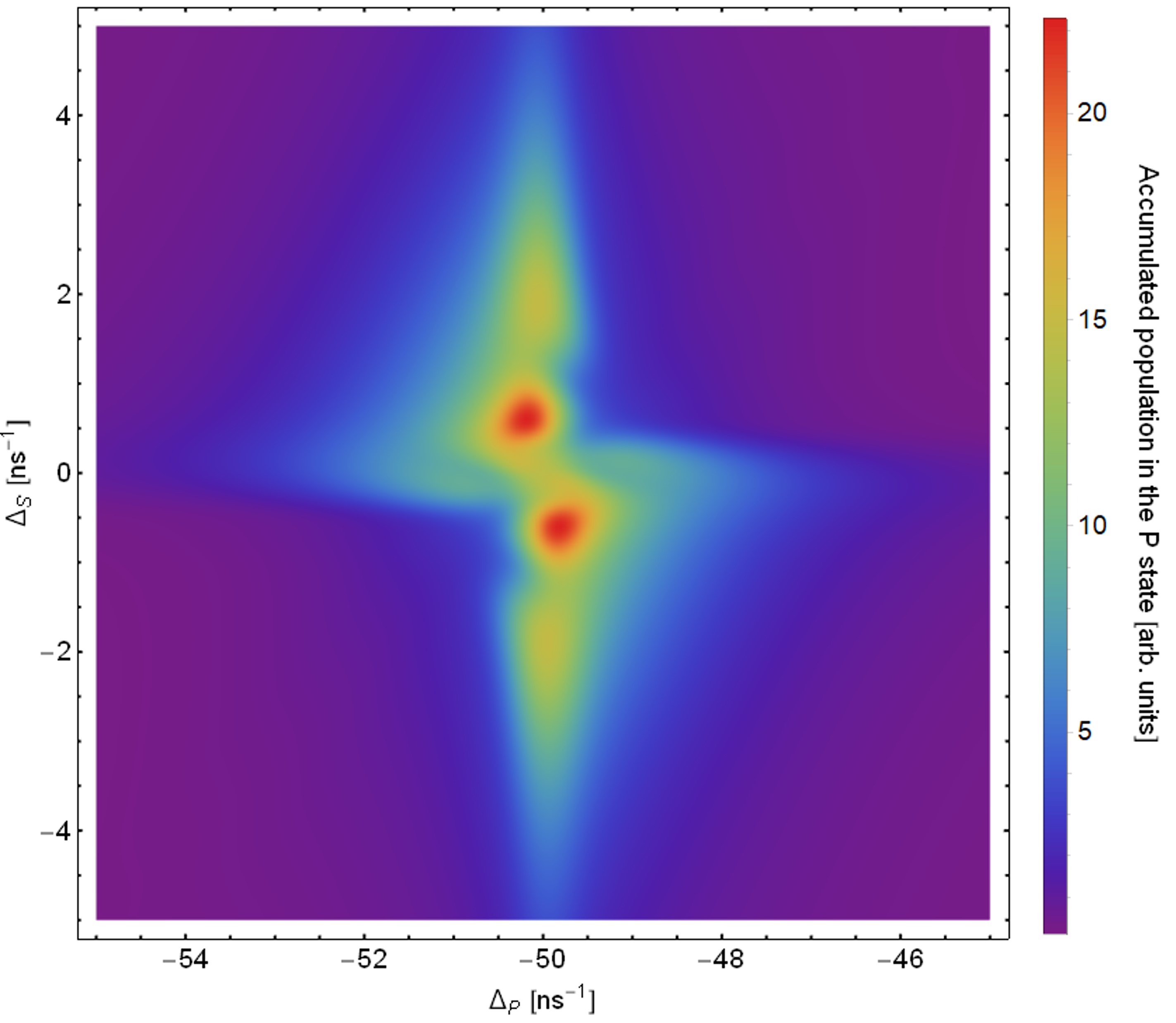}
\caption{\label{Simu4} Accumulated population in the \P state after 200 repump cycles as a function of the detunings of the Pump and Stokes lasers. The laser pulses are coincident, with a pulse duration of 10\,ns, and the Rabi frequencies are $\Omega_P=3.5\,$ns$^{-1}$ for the Pump laser and $\Omega_S=2.8\,$ns$^{-1}$ for the Stokes laser.}
\end{center}
\end{figure}

Figure,\ref{Simu4} shows the accumulated population in the \P state after 200 repump cycles as a function of the Pump and Stokes laser detunings. Although there is a certain tolerance in the detunings, achieving successful experimental implementation requires strict control over any potential source that could shift the absorption frequencies. In this regard, ensuring tunability of the laser system is essential, even though this is typically an inherent characteristic of nanosecond lasers. Such tunability would allow precise adjustments to compensate for any drifts or shifts in the transition frequency, thereby maintaining optimal conditions for the experiment. 

\section{Alternative laser scheme for the implementation of BaTa} 

\begin{figure}[ht!]
\begin{center}
\includegraphics[width=8.3cm, height=5cm]{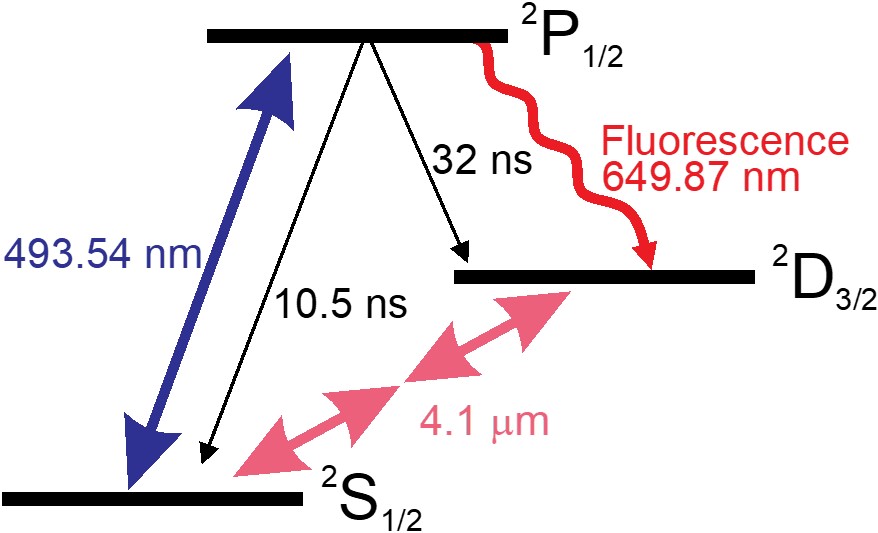}
\caption{\label{levelscheme2} Alternative laser scheme for the implementation of Ba-tagging (BaTa). In this approach, the repumping of the population from the metastable \D state is performed via the \D$\rightarrow$\S two-photon transition instead of the single-photon \D$\rightarrow$\P transition.}
\end{center}
\end{figure}

Considering the energy-level structure of barium ions, an alternative laser scheme could be proposed for the pump/repump cycles to address the population trapped in the metastable \D state. Instead of repumping via the \P state, which presents the drawback of the overlapping between the repumping laser wavelength and the desired detection fluorescence (requiring pulsed fluorescence detection to separate signals), it is possible to repump the population through the ground state \S using a two-photon transition (see Fig.,\ref{levelscheme2}). This alternative configuration offers a significant advantage: it enables continuous fluorescence detection, as the detection wavelength from the \P$\rightarrow$\D transition is different from the Pump and Stoke laser wavelengths. This fact allows for more straightforward data acquisition without temporal gating, potentially simplifying some aspects of the experimental setup.

However, this approach also introduces notable challenges that, in principle, make it less practical for experimental implementation. Firstly, the two-photon transition required for repumping the \D state  population via the \S state necessitates a laser operating at 4.1\,$\mu$m. Achieving such a wavelength with long-pulse laser systems is technologically demanding. Mid-infrared lasers, particularly those with nanosecond pulse durations, are less accessible. Additionally, since this radiation is invisible to the naked eye, ensuring an accurate beams positioning and overlap becomes an arduous task. Secondly, the requirement for a two-photon transition adds an additional difficulty. Transitions of higher order, i.e., those involving more than one photon, demand significantly higher laser intensities, scaling the transition probability linearly with the laser power. This imposes additional requirements on the laser systems and necessitates exceptional control over experimental parameters, such as laser beam quality, spatial overlap, and frequency stability, to ensure efficient repumping. Such precision becomes a limiting factor, particularly in experiments already constrained by technical difficulties.

Thus, we can conclude that, while the laser scheme illustrated in Fig.,\ref{levelscheme2} provides an interesting alternative that may allow continuous fluorescence measurement, it does not present, in our opinion, a clear experimental advantage over the approaches discussed in Fig.,\ref{levelscheme}. The challenges associated with mid-infrared laser generation and the high-intensity requirements for two-photon transitions outweigh the potential benefits in this scenario.

\section{Conclusions}

In this manuscript, we have described a potential implementation of BaTa for the @NEXT experiment, relying solely on laser-matter interactions and the structure of the barium ion. While the experimental implementation is far from straightforward, as it requires a combination of lasers due to the level structure of barium, which includes a metastable state, the experimental conditions are highly demanding, and the detection requires extreme precision. However, based on the simulations discussed, the implementation of the technique is feasible. One advantage of the proposed BaTa implementation is that it does not introduce contaminants into the experimental chamber, which could compromise the detection of the double beta transition. Providing access to all positions within the chamber, for instance, via windows, and employing an appropriate servo system to direct the lasers to the origin point of the double beta decay should be sufficient for implementation. The application of the BaTa technique would significantly enhance the NEXT experiment and contribute to answering the questions that still persist about the neutrinos. 

\section{Acknowlegments}
The author thanks the members of the NEXT collaboration for the most fruitful discussions. 
\appendix
\section{The incoherent limit}\label{AppIncohLimit}

In a situation where the pulse duration is much longer than the radiative decay times, it can be shown that the population dynamics can be effectively described in terms of rate equations rather than in terms of the time dependent Schr\"odinger equation, which considerable simplifies the analysis \cite{Sh90}.
\begin{equation}
\begin{array}{l}
\label{rate}
\dfrac {\mathrm{d} P_1}{\mathrm{d} t}=-B_{12}\frac{I_P(t)}{c}P_1(t)+(B_{21}\frac{I_P(t)}{c}+A_{21})P_2(t)\nonumber \\ \\
\dfrac {\mathrm{d} P_2}{\mathrm{d} t}=B_{12}\frac{I_P(t)}{c}P_1(t)-(B_{21}\frac{I_P(t)}{c}+A_{21})P_2(t)-(B_{23}\frac{I_S(t)}{c}+A_{23})P_2(t)+B_{32}\frac{I_S(t)}{c}P_3(t)\nonumber  \\ \\
\dfrac {\mathrm{d} P_3}{\mathrm{d} t}=(B_{23}\frac{I_S(t)}{c}+A_{23})P_2(t)-B_{32}\frac{I_S(t)}{c}P_3(t) \nonumber
\end{array}
\end{equation}
where $P_i$ represent the populations, c denotes the speed of light, $I_P$ and $I_S$ denote the laser intensities for Pump and Stokes laser respectively, and A and B the Einstein coefficients for spontaneous emission and stimulated absorption with the relation $B_{ij}=B_{ji}$. 

In the central part of the laser pulse, the population is almost stationary. Thus, we can consider that $\dot{P_i}\rightarrow0$. With this approximation the population for every state reads: 

At the central part of the laser pulses, the population of each state reaches a near-stationary condition. Therefore, we can assume $\dot{P_i}\rightarrow0$ allowing us to simplify the rate equations by treating the population as constant. Under this approximation, the population of the state \P and \D as a function of the state \P can be expressed as:
\begin{equation}         
P_2=\frac{B_{12}I_PP_1}{A_{21}c+B_{12}I_P} \qquad
P_3=\frac{B_{12} I_P (A_{23} c + B_{23} I_S) P_1}{B_{23} (A_{21} c + B_{12} I_P) I_S}.
\end{equation}
Since our interest is the population difference between state \P and states \S and \D, which directly correlates with the fluorescence signal, we define the following function $\vartheta$:
\begin{equation}
\vartheta=\frac{P_2-(P_1+P_3)}{P_1+P_2+P_3}=-\frac{A_{23} B_{12} c I_P + A_{21} B_{23} c I_S + B_{12} B_{23} I_P I_S}{
 A_{23} B_{12} c I_P + A_{21} B_{23} c I_S + 3 B_{12} B_{23} I_P I_S}
\end{equation}
The defined function $\vartheta$ grows as a function of the laser intensities until a saturation value of $\vartheta=-\frac{1}{3}$, indicating the maximum achievable population difference between state \P and states \S and \D. Also, the direction of growth of $\vartheta$ is defined by the gradient: 
\begin{equation}
\nabla \vartheta=\left(\frac{\partial \vartheta}{\partial I_P}, \frac{\partial \vartheta}{\partial I_S}\right)=\left(            
\frac{2 A_{21} B_{12} B_{23}^2 c I_S^2}{(A_{23} B_{12} c I_P + B_{23} (A_{21} c + 3 B_{12} I_P) I_S)^2}, 
\frac{2 A_{23} B_{12}^2 B_{23} c I_P^2}{(A_{23} B_{12} c I_P + B_{23} (A_{21} c + 3 B_{12} I_P) I_S)^2}
\right).
\end{equation}
According to this expression, the direction of $\vartheta$ increase does not exactly follow the direction $I_P=I_S$ but slightly deviates due to the difference in the stimulated and spontaneous emission coefficient of the transition \S$\rightarrow$\P and \P$\rightarrow$\D. This explains the deviation observed in Fig.\,\ref{Simu2}.


\begin{thebibliography}{7}

\bibitem{Athar22} M. Sajjad Athar, et al., Prog. Part. Nucl. Phys., 124, 103947, 2022. 

\bibitem{Abe23} K. Abe, et al., Eur. Phys. J. C 83, 782, 2023.

\bibitem{Acero22} M. A. Acero, et al., Phys. Rev. D 106, 032004, 2022. 

\bibitem{Abed21} A. Abed Abud, et al., Instruments, 5(4), 31, 2021. 

\bibitem{Yoshimura15} M. Yoshimura, N. Sasao, and M. Tanaka, Prog.Theor. Exp. Phys., 5, 053B06, 2015.

\bibitem{Agostini20} M. Agostini, et al., Phys. Rev. Lett. 125, 252502, 2020. 

\bibitem{Albert16} J. Albert, et al.,Phys. Rev. C 93. 035501, 2016.

\bibitem{Abe23} S. Abe, et al., Phys. Rev. Lett. 130, 051801, 2023. 

\bibitem{Novella23} P. Novella, et al., J. High Energ. Phys. 2023, 190, 2023.

\bibitem{Gando13} A. Gando, et al., Phys. Rev. Lett. 110(6), 062502, 2013.

\bibitem{Gando16} A. Gando, et al., Phys. Rev. Lett. 117, 082503–082506, 2016.

\bibitem{Majorana37} E. Majorana, Nuovo Cim. 14, 171–184, 1937.

\bibitem{Sakharov67} A. D. Sakharov, Pis’ma Z. Eksp. Teor. Fiz. 5, 32–35, 1967.

\bibitem{Fukugita86} M. Fukugita and T. Yanagida, Phys. Lett. B 174, 45–47, 1986.

\bibitem{Mohapatra80} R. N. Mohapatra and G. Senjanovic, Phys. Rev. Lett. 44, 912–915, 1980.

\bibitem{Elliot02} S. R. Elliott and P. Vogel, Annu. Rev. Nucl. Part. Sci. 52, 115–151, 2002.

\bibitem{Moe91} M. K. Moe, Phys. Rev. C 44, 931–934, 1991.

\bibitem{Danilov00} M. Danilov, Phys. Lett. B 480, 12–18, 2000. 

\bibitem{Sinclair11} D. Sinclair, et al., J. Phys. Conf. Ser. 309, 012005, 2011.

\bibitem{Mong15} B. Mong, et al., Phys. Rev. A 91, 022505–022513, 2015.

\bibitem{nEXO19} nEXO Collaboration, Nature 569, 203–207, 2019.

\bibitem{Albert15} J. B. Albert, et al, Phys. Rev. C 92, 045504–045510, 2015.

\bibitem{Nygren15} D. R. Nygren, J. Phys. Conf. Ser. 650, 012002, 2015. 

\bibitem{Thapa19} P. Thapa, et al., Sci. Rep. 9, 15097, 2019. 

\bibitem{Rivilla20} I. Rivilla, et al., Nature 583, 48–54, 2020.

\bibitem{Haefner24} J. Haefner, et al., Eur. Phys. J. C 84, 518, 2024.

\bibitem{Sanso10} J. E. Sansonetti and J. J. Curry, J. Phys. Chem. 39, 4, 2010.

\bibitem{Abde13} K. Addessalem, et al., J. Phys. Chem A 117, 8925-8938, 2013.

\bibitem{Curry04} J. Curry, Compilation of wavelengths, energy levels, and transition probabilities for Ba I and Ba II, Journal of Physical and Chemical Reference Data, 2004.

\bibitem{Sh90}
B. W. Shore, The Theory of Coherent Atomic Excitation, Wiley, NY, 1990.

\bibitem{Cohen}
 C. Cohen-Tannoudji, B. Diu, and F. Lao\"e, Quantum Mechanics Vol. 1 and Vol. 2, Wiley-Vch, 2005.
 
\bibitem{Hilborn}
R. C. Hilborn Am. J. Phys. 50, 982–986, 1982.

\bibitem{Shore08}
B. W. Shore, Acta Phys. Slov., 58, 243–486, 2008.

\bibitem{Demtroder03} W. Demtr\"{o}der, Laser Spectroscopy: Basic Concepts and Instrumentation. Springer, 2003.

\bibitem{Martin16} G. Martín-Albo et al., J. High Ener. Phys., 2016, 5, 159, 2016.
%

%




\end{thebibliography}
\end{document}